# The status of the NIR arm of the SOXS Instrument toward the PAE

Fabrizio Vitali[1], Matteo Genoni[2], Matteo Aliverti[2], Kalyan Radhakrishnan[3], Federico Battaini[3], Paolo D'Avanzo[2], Francesco D'Alessio[1], Giorgio Pariani[2], Luca Oggioni[2], Salvatore Scuderi[4], Davide Ricci[3], Eugenio Martinetti[5], Antonio Miccichè[5], Gaetano Nicotra[5], Mirko Colapietro[6], Sergio D'Orsi[6], Matteo Munari[3], Luigi Lessio[3], Simone Di Filippo[3], Andrea Scaudo[2], Giancarlo Bellassai[3], Rosario Di Benedetto[5], Giovanni Occhipinti[5], Marco Landoni[2], Matteo Accardo[7], Leander Mehrgan[7], Derek Ives[7], Carlotta Scirè[5], Sergio Campana[2], Pietro Schipani[6], Riccardo Claudi[3], Giulio Capasso[6], Marco Riva[2], Ricardo Zanmar Sanchez[6], José Antonio Araiza-Durán[8], Iair Arcavi[9], Andrea Baruffolo[3], Sagi Ben-Ami[10], Anna Brucalassi[8], Rachel Bruch[10], Enrico Cappellaro[3], Rosario Cosentino[11], Marco De Pascale[3], Massimo Della Valle[6], Avishay Gal-Yam[10], Marcos Hernandez Díaz[10], Ofir Hershko[10], Jari Kotilainen[12], Hanindyo Kuncarayakti[13], Gianluca Li Causi[1], Laurent Marty[6], Seppo Mattila[13], Hector Pérez Ventura[11], Giuliano Pignata[14], Michael Rappaport[10], Adam Rubin[7], Bernardo Salasnich[3], Stephen Smartt[15], Maximilian Stritzinger[16], David Young[15]

[1]INAF Osservatorio Astronomico di Roma, [2]INAF - Osservatorio Astronomico di Brera, [3]INAF - Osservatorio Astronomico di Padova, [4]INAF - Istituto di Astrofisica Spaziale e Fisica Cosmica, [5]INAF - Osservatorio Astrofisico di Catania, [6]INAF - Osservatorio Astronomico di Capodimonte, [7]European Southern Observatory, [8]INAF-Osservatorio Astrofisico di Arcetri, [9]Tel Aviv University, [10]Weizmann Institute of Science , [11]INAF - Fundación Galileo Galilei, [12]FINCA - Finnish Centre for Astronomy with ESO, [13]Tuorla Observatory, Department of Physics and Astronomy, University of Turku, [14]Universidad Andres Bello, [15]Queen's University Belfast, School of Mathematics and Physics, [16]Aarhus University

**ABSTRACT**

The Son Of X-Shooter (SOXS) is a single object spectrograph, built by an international consortium for the 3.58-m ESO New Technology Telescope at the La Silla Observatory [1]. It offers a simultaneous spectral coverage over 350-2000 nm, with two separate spectrographs. In this paper we present the status of the Near InfraRed (NIR) cryogenic echelle cross-dispersed spectrograph [1], in the range 0.80-2.00 μm with 15 orders, equipped with an 2k x 2k Hawaii H2RG IR array from Teledyne, working at 40K, that is currently assembled and tested on the SOXS instrument, in the premises of INAF in Padova. We describe the different tests and results of the cryo, vacuum, opto-mechanics and detector subsystems that finally will be part of the PAE by ESO.

*fabrizio.vitali@inaf.it, +39 06 94286462

# 1. INTRODUCTION

The Son Of X-Shooter (SOXS) is a single object spectrograph, built by an international consortium for the 3.58-m ESO New Technology Telescope at the La Silla Observatory. It offers a simultaneous spectral coverage over 350-2000 nm, with two separate spectrographs. The Near InfraRed (NIR) arm consists in a fully criogenic echelle dispersed spectrograph, working in the range 0.800-2.005 mm, the optical design is based on a 4C echelle and the dispersion is obtained via a main disperser grating and three cross-disperser in double-pass. The NIR spectrum is dispersed in 15 orders and imaged on an Hawaii H2RG IR array from Teledyne, controlled via the new NGC controller from ESO. The spectrograph is cooled down via a Leybold M10 system. All the subsystems (cryo, vacuum, opto-mechanics and detector) have been assembled and tested at the INAF- OAB premises in Merate and then the NIR arm has been moved to the INAF-OAPd premises in Padova, to be assembled on the SOXS flange and tested for the PAE (Preliminary Acceptance in Europe), before the final shipping to Chile.

# 2. TOWARDS THE PAE

After the assembly and test phase in the premises of INAF-OAB in Merate [3], the NIR cryostat was moved to the INAF-OAP premises in Padova, where a large laboratory hosts an NTT Nasmyth derotator simulator, lent by ESO, on which SOXS has been assembled and it is being tested towards the PAE.

The Nasmyth simulator actually hosts the whole SOXS instrument [4], including all the subsystems, e.g., the UV-VIS and NIR spectrographs, the Common Path, the Acquisition Camera and the Calibration box (Figure 1). The derotation system has been installed and tested as well: it hosts and derotates all the SOXS cables, the He pipes (for the NIR cryocooler system) and the $LN_2$ pipe, for the UV-VIS CFC cooling system. To complete the instrument installation, the two electronics racks are set as in the final configuration in the Nasmyth room of the NTT telescope, in La Silla (Chile).

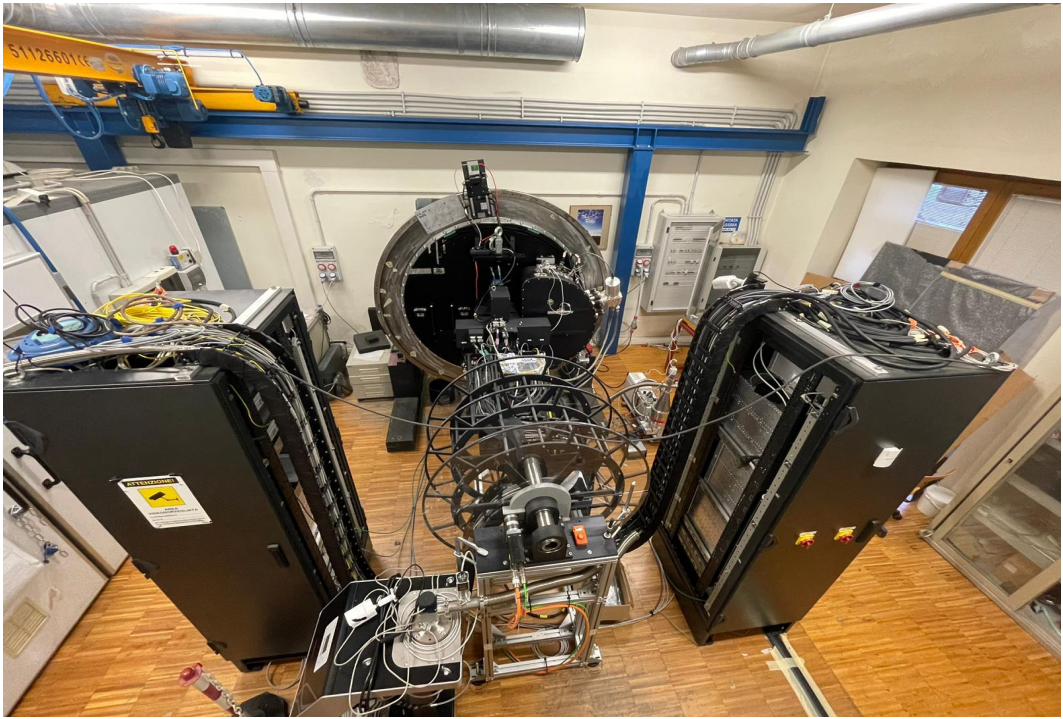

Figure 1: the SOXS Instrument assembled in the laboratory at INAF-OAP, in Padova.

The NIR cryostat was firstly mounted on its tilting table (Figure 2, left), to perform a first run of tests, devoted to check the opto-mechanics stability after the shipping. Then, the cryostat was opened for the final mounting and checks, in particular to check all the baffle and cover inside the cryostat, to minimize the spurious thermal light and minimize the instrumental dark current (Figure 2, right). Moreover, a new cryocooler has been mounted, to substitute the older and less efficient version.

Then, the NIR cryostat was mounted on the SOXS flange, through three cinematic mounts, and currently it is part of the whole SOXS instrument.

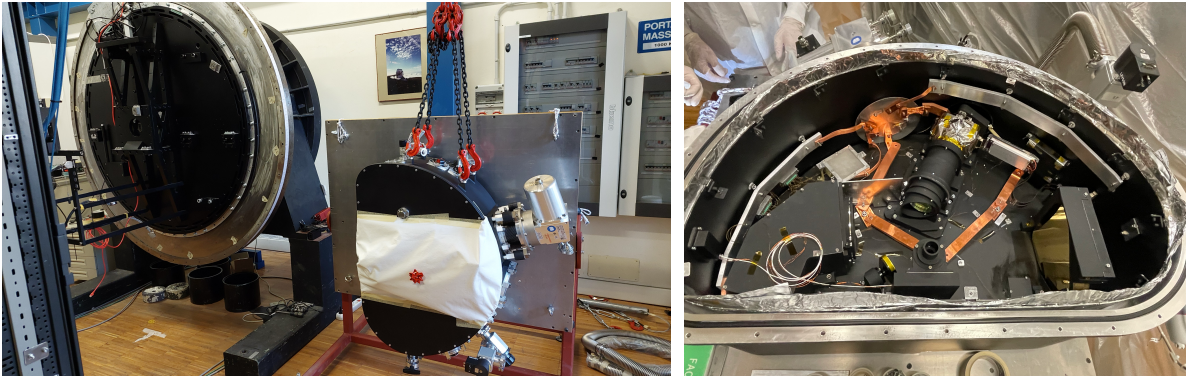

*Figure 2: the SOXS NIR cryostat mounted on the tilting table (left) and opened for the final check on the baffle system (right).*

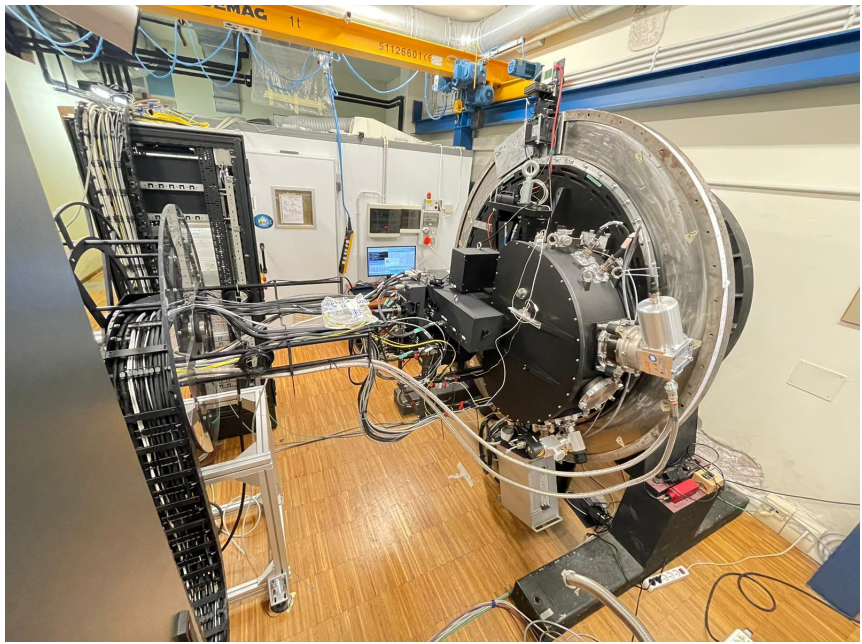

*Figure 3: the SOXS-NIR cryostat mounted on the flange of SOXS.*

**Optical tests**

The tests previously performed in Merate [3] have confirmed that the optics fulfil the requirements. In Padova we repeated all the optical tests, to check possible disalignment due to the shipping.

The optical quality has been checked, after the shipment and the mounting on the SOXS flange, with a CP simulator applied on the cover of the NIR cryostat. The results, in terms of FWHM, are shown in Table 1 and Figure 4: the FWHMs of 25 lines, homogeneously chosen along the echellogram. are in line with the tests performed in Merate. The flexure test done is a comparison of the 25 spectral features X, Y positions. The retrieved X and Y positions in Padova frames are stable within 1.5 pixels w.r.t. the ones taken in Merate (Figure 5), which resulted in agreement with the requirements.

*Table 1: the comparison of the optical test in Merate and in Padova.*

|  | FWHM X (main-disp) |  | FWHM Y (cross-disp) |  |
| --- | --- | --- | --- | --- |
| Date | Mean | Median | Mean | Median |
| 19.12.2023 | 2.0295 | 1.9382 | 2.1006 | 2.0481 |
| 11.03.2024 | 2.0664 | 1.9547 | 2.0443 | 1.9081 |

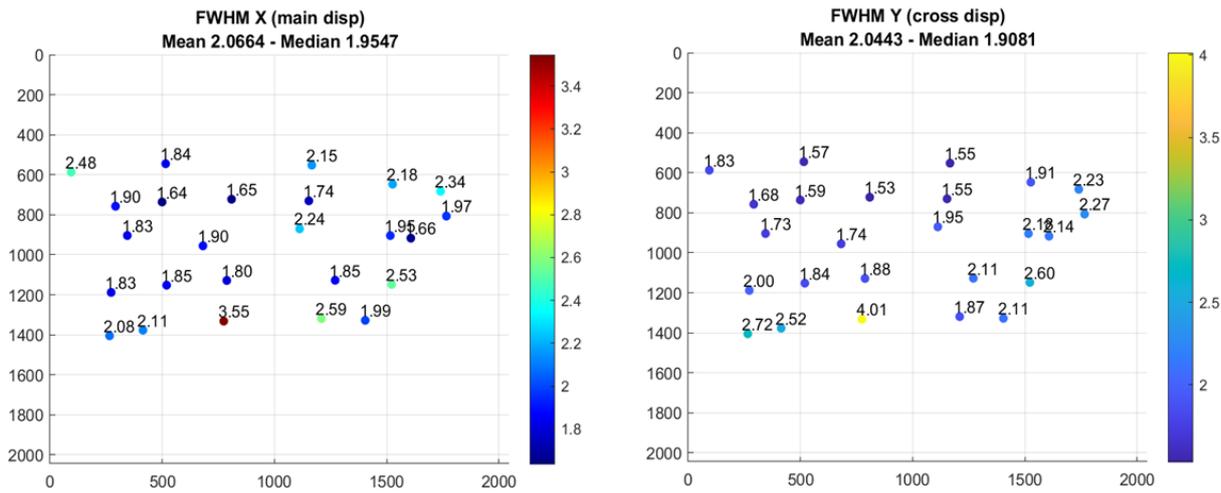

*Figure 4: the FWHMs of 25 lines, homogeneously chosen along the echellogram.*

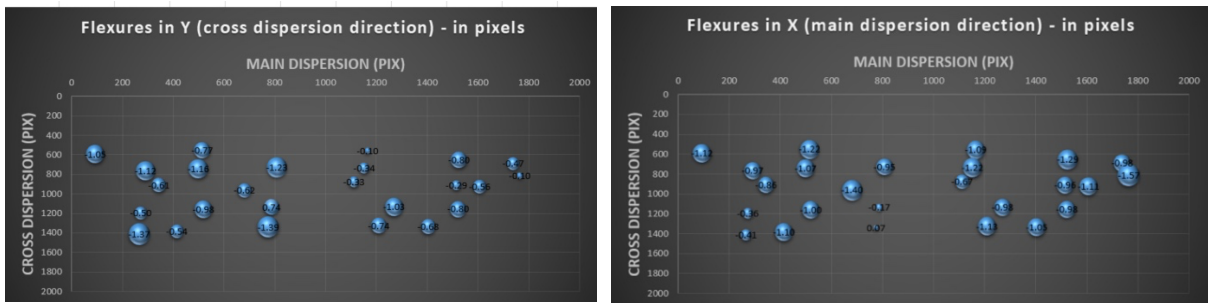

*Figure 5: the flexures test results, taken on the same 25 spectral lines analyzed for the optical quality check.*

From the same set of spectral lines, we computed the Resolving Power, R, as $R = \lambda/\delta\lambda$, where $\delta\lambda$ is computed from the FWHM-X value computed for the 25 wavelengths taking into account their linear dispersion. In Figure 6 R is always greater than the required resolution power of 3500.

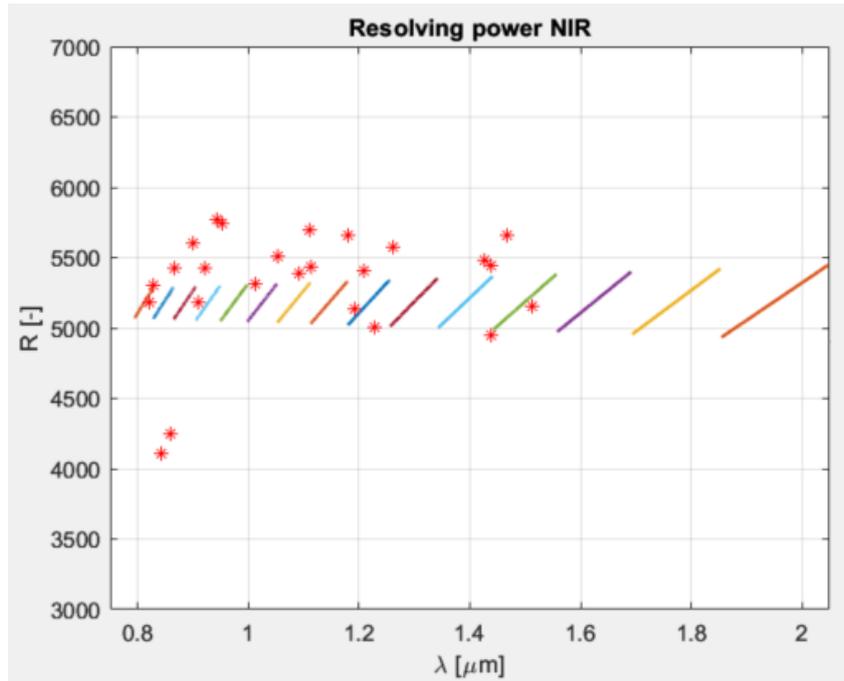

*Figure 6: Resolving Power plots of the 1 arcsec slit, related to FWHM in main dispersion computed in Figure 4.*

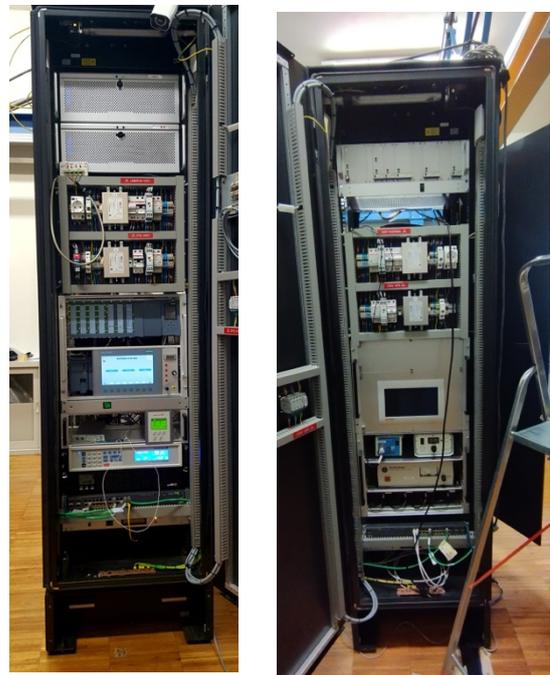

### The Control Electronics

It manages all the SOXS motorized functions is hosted in two distinct cabinets, the NIR slit exchanger controller and the NIR NGC Power supply are located on the rear side of the cabinet in a 19" subrack.

In the second cabinet, all the cryo-vacuum control electronics are hosted: the Siemens PLC, the vacuum gauges controller and the Lakeshore temperature controller for the NIR and VIS detectors. The NIR NGC LLCU is hosted as well, on the top of this cabinet [5].

### The Instrument Control Software

SOXS is controlled vis the Instrument Control System [ICS], a series of GUIs (Figure 7). The first one is the custom OS_CONTROL panel, which is responsible to control the status of each subsystem of SOXS, including the NIR spectrograph. The second is the custom ICS panel, an engineering panel for the subsystem that controls motors and monitors all SOXS sensors. In the selected TAB, all the monitored NIR sensors are shown. The third is the NGCIR panel, which is not custom and is part of the NGC software system provided by ESO. It controls the NIR spectrograph [6].

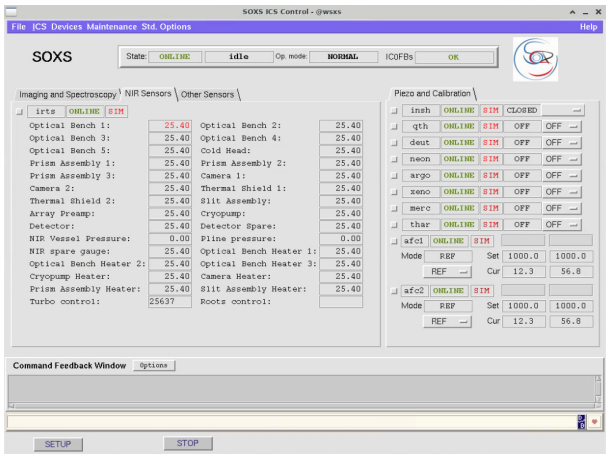
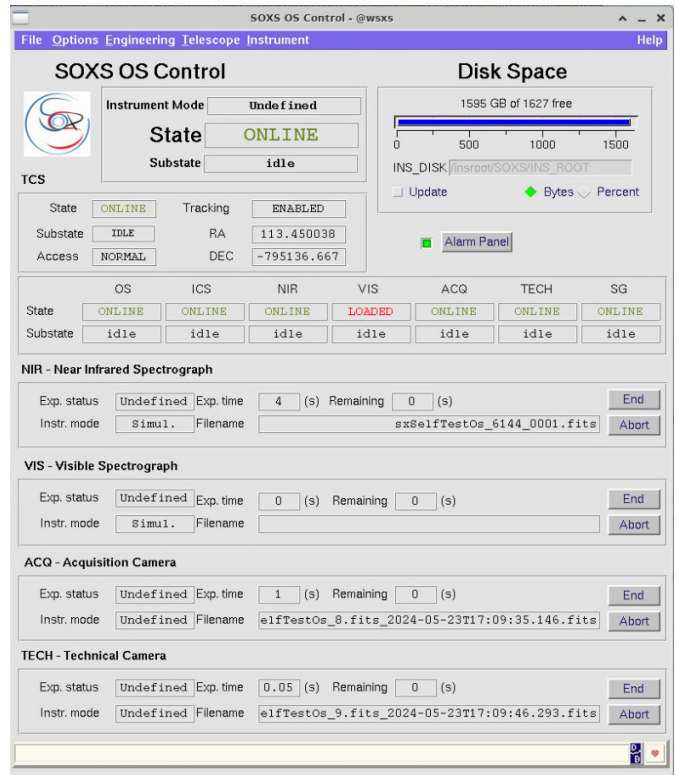
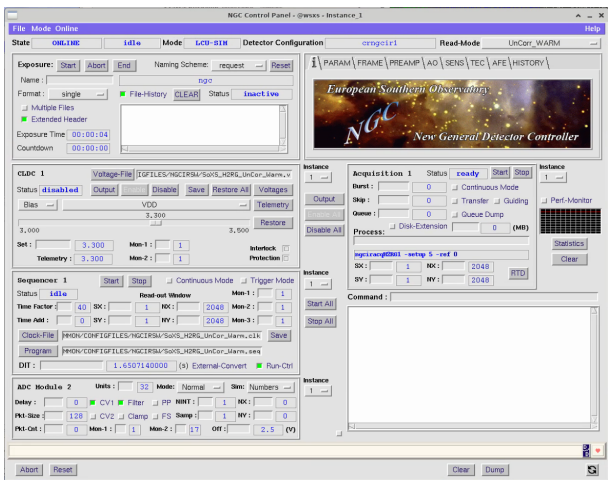

*Figure 7: the GUIs of the ICS.*

**The cryogenics performances**

The NIR cryostat is cooled through the Leybold COOLPOWER 250 MDi cryocooler, that can cool down the detector at 40 K in about 45 h, and the whole cryostat at 145 K in about 60 h. In Figure 8 we show the main cooling curves, which refers to the cryostat bench (145 K), the H2RG preamplifier (120 K), the cryo-pump (50 K), the cryocooler cold head (< 25 K) and the H2RG detector (40 K). During the cool down, the cooling rate of the detector was always greater than 0.25 K/min, then in perfect safe condition.

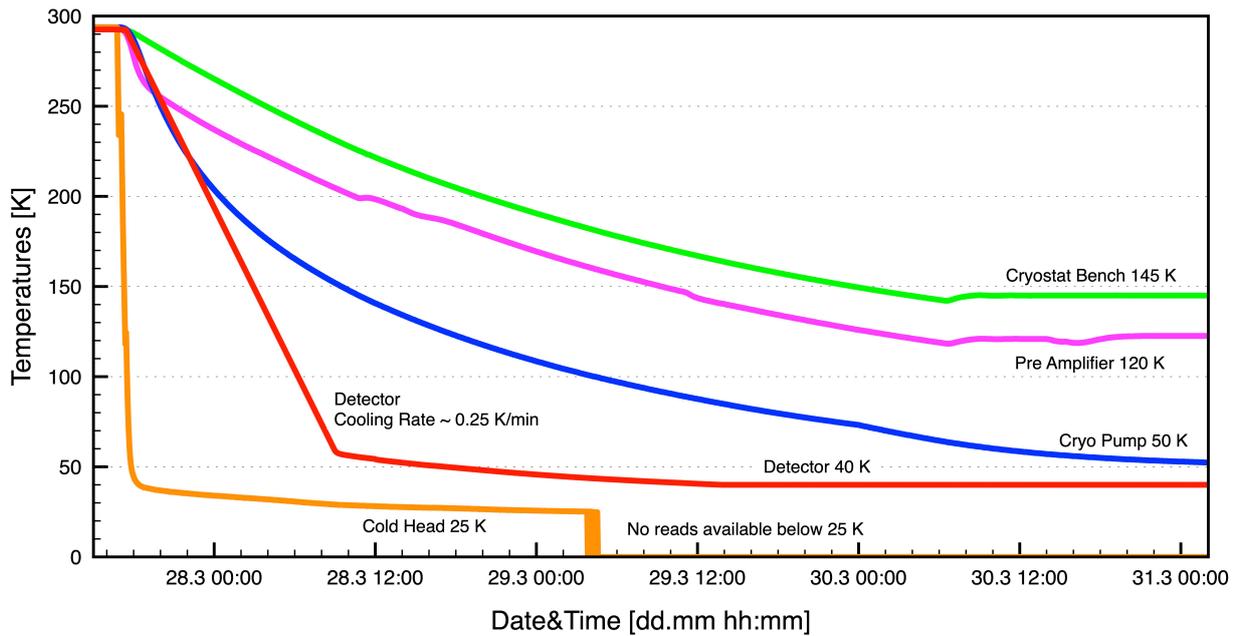

*Figure 8: the cooling curves for the NIR cryostat of SOXS.*

In Padova, we test again the performances of the H2RG detector, in terms of gain, dark current and linearity. The gain has been confirmed at 2.3 e-/DN. In Figure 9 (left) you can see the measured DC average of 0.011 DN/s/pixel, resulting in 0.025e-/s/pixel.

NIR detector linearity (Figure 9, right) has been measured taking spectra of an halogen (QTH) lamp with the 5" slit. The saturation occurs at ~ $5 \times 10^4$ counts, the linearity starts to get lost ~ $3 \times 10^4$ counts (same results as in Merate).

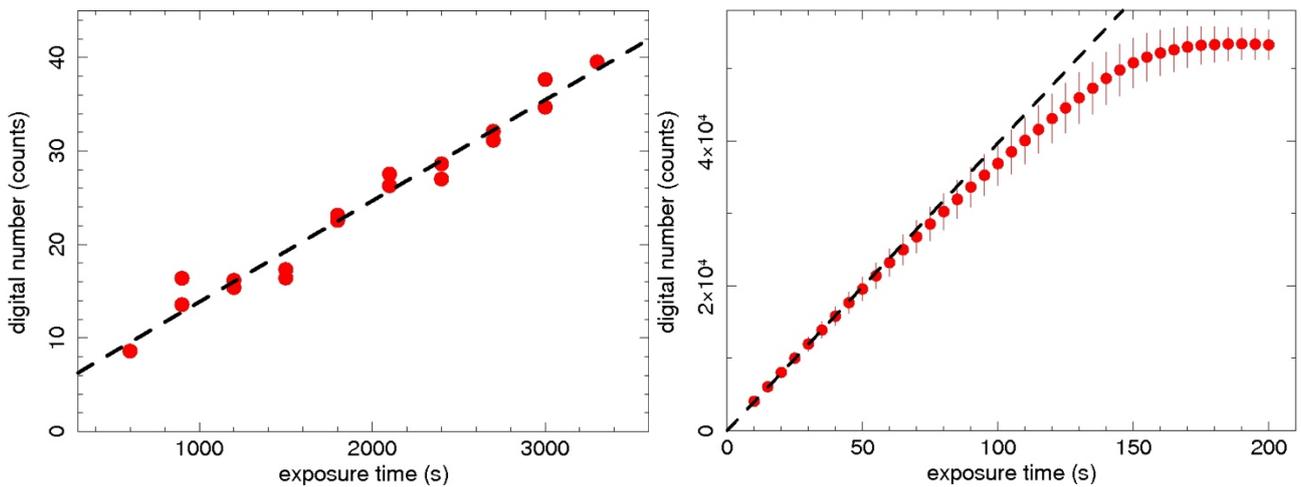

*Figure 9: the measured dark current and the linearity curve.*

## 3. CONCLUSIONS

We are currently in the Preliminary Acceptance in Europe (PAE) phase of the SOXS instrument in the premises of the INAF-OAP in Padova. At the time of writing, all the subsystem of the NIR arm of SOXS are working well, we are currently fine tuning all the sw and hw procedures and check all the working parameters and detector performances. The PAE will last few months and we expect to ship the whole instrument to the ESO La Silla observatory, in Chile, for the end of this year.